\documentclass[a4paper]{article}

\usepackage[english]{babel}
\usepackage[utf8x]{inputenc}
\usepackage{amsmath}
\usepackage{amsopn}
\usepackage{amstext}
\usepackage{amssymb}
\usepackage{amscd}
\usepackage{graphicx}
\usepackage{epsfig}
\usepackage{listings}
\usepackage{color}
\usepackage[colorinlistoftodos]{todonotes}
\usepackage{hyperref}

\providecommand{\abs}[1]{\lvert#1\rvert}
\providecommand{\norm}[1]{\lVert#1\rVert}
\newcommand{\bfb}{\mathbf{b}}
\newcommand{\bfr}{\mathbf{r}}
\newcommand{\bfs}{\mathbf{s}}
\newcommand{\bft}{\mathbf{t}}
\newcommand{\bfv}{\mathbf{v}}
\newcommand{\bfw}{\mathbf{w}}
\newcommand{\bfx}{\mathbf{x}}
\newcommand{\bfy}{\mathbf{y}}
\newcommand{\bfz}{\mathbf{z}}
\newcommand{\bfA}{\mathbf{A}}
\newcommand{\bfI}{\mathbf{I}}
\newcommand{\bfX}{\mathbf{X}}
\newcommand{\bfR}{\mathbf{R}}

\newcommand{\R}{\mathbb{R}}

\DeclareMathOperator{\sig2}{\sigma^2_{max}}

\DeclareMathOperator{\E}{E}
\DeclareMathOperator{\Var}{Variance\;}
\DeclareMathOperator{\T}{T}
\newcommand{\st}{\mbox{subject to\;}}
\newcommand{\Q}{\mathcal{Q}}
\newcommand{\Qr}{\mathcal{Q}_r}

\title{Regression techniques for Portfolio Optimisation using MOSEK}
\author{Thomas Schmelzer\footnote{Z\"urcherstr. 2, Altendorf, 8852, Switzerland, thomas.schmelzer@gmail.com}, Raphael Hauser\footnote{Mathematical Institute, University of Oxford, Andrew Wiles Building, Radcliffe Observatory Quarter, Woodstock Road, Oxford, OX2 6GG, United Kingdom, hauser@maths.ox.ac.uk},\\Erling D.~Andersen\footnote{MOSEK ApS, Fruebjergvej 3, Box 16, Copenhagen, 2100, Denmark, support@mosek.com}, Joachim Dahl$^\ddag$}

\begin{document}

\definecolor{keywords}{RGB}{255,0,90}
\definecolor{comments}{RGB}{0,0,113}
\definecolor{red}{RGB}{160,0,0}
\definecolor{green}{RGB}{0,150,0}
 
\lstset{language=python, 
        basicstyle=\ttfamily\small, 
        commentstyle=\color{comments},
        stringstyle=\color{red},
        showstringspaces=false,
        frame=tblr,                         
        numbers=left,
        numberstyle=\footnotesize,
        numbersep=1em,
        }
 
\maketitle

\begin{abstract}
Regression is widely used by practioners across many disciplines. 
We reformulate the underlying optimisation problem as a second-order conic program providing the flexibility often needed in applications.
Using examples from portfolio management and quantitative trading we solve regression problems with and without constraints. Several Python code fragments are given\footnote{Code and data are available online at \url{http://www.github.com/tschm/MosekRegression}}.
\end{abstract}

\section{Introduction}
Regression is the hammer in the toolbox for any professional quant. It is widely used and sheer force can often yield amazing results. 
However, regression goes far beyond the simple concept of fitting a line into a cloud of points.

Regression is closely related to conic programming and while keeping the mathematical details at a minimum we discuss this connection in Section $2$. 
This makes regression a versatile tool for portfolio optimisation as we are able to apply constraints and bounds. 

In Section $3$ we address the closely related regularisation problem. 
In the context of portfolio optimisation regularisation terms mimic trading costs. Such penalties help to tame the underlying optimisation problem.

In Section $4$ we discuss typical problems in the management of equity portfolios. We discuss some common practical concepts and their implementation as a conic program.

We sketch how to generate data-driven estimators for future returns in Section $5$.

Finally we demonstrate using real world data how to construct common portfolios using the tools and concepts introduced in previous sections.


\section{Regression}\label{sec:regression}
The core regression problem is to model the linear relationship 
between $m$ explanatory variables $\bfX \in \mathbb{R}^{n \times m}$ 
and the dependent variable $\bfy \in \mathbb{R}^n$. 
The columns of $\bfX$ are the explanatory variables $\bfx_1, \bfx_2, \ldots, \bfx_m$.
We find coefficients $w_1, w_2, \ldots, w_m$ such that the weighted sum
\[
\bfX \bfw = \sum_{i=1}^m \bfx_i w_i 
\]
has minimal Euclidean distance to $y$,
\begin{equation}\label{leastSquare}
\min_{\bfw \in \R^m}\; \norm{\bfX\bfw - \bfy}_2.
\end{equation}
The term $\bfr=\bfX \bfw - \bfy$ is the \emph{residual}. 
The $2$-norm of the residual is $\norm{\bfr}_2 = \sqrt{\bfr^T \bfr} = \sqrt{\sum_{i=1}^n r_i^2}$. Note that the $2$-norm resembles a scaled standard deviation of the vector $\bfr$ if this vector is centered (mean zero).

Equation (\ref{leastSquare}) is an \emph{unconstrained least squares problem} as we are minimizing (the square root of) the sum of squared residuals.

It is more common in literature to minimise the square of the $2$-norm,
\begin{equation}\label{leastSquare2}
\min_{\bfw \in \R^m}\;  \norm{\bfX \bfw - \bfy}^2_2.
\end{equation}
Obviously, the problems in Equation (\ref{leastSquare}) and Equation (\ref{leastSquare2}) have the same solution.
Throughout this article we assume $\bfX$ has more rows than columns. We call such systems overdetermined. Underdetermined systems are rare in practical applications in finance and require the application of techniques introduced in Section $3$ to make their solution unique.  

\subsection{The normal equations}
The residual has to be orthogonal to the range of $\bfX$. 
This geometric insight is the base for the most powerful algorithms 
but also the core idea underlying the normal equations revealing an explicit solution for this unconstrained problem
\[
\bfX^{\T} \bfX \bfw = \bfX^{\T} \bfy.
\]
Please avoid solving those equations directly or even worse by computing the inverse of $\bfX^{\T} \bfX$ explicitly. 
This can go terribly wrong, in particular if $\bfX$ has almost linear dependent columns.
Leaving those numerical reasons aside there is no elegant way to state bounds and constraints on $\bfw$ using this approach. Note that solving those equations and then applying constraints by modifying the unconstrained solution leads to very suboptimal results in many cases.

\subsection{Cones}
The desired flexibility is achieved by embedding regression in a more general and powerful concept known as conic programming. For the purpose of this paper it is enough to understand quadratic and rotated quadratic cones.

\begin{figure}
\centering
\def\svgwidth{0.35\columnwidth}
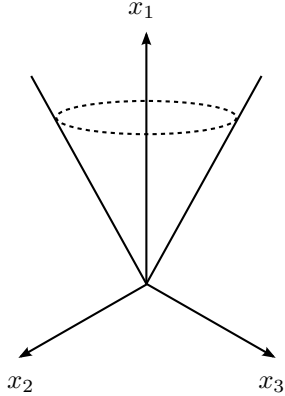
\caption{A quadratic or second-order cone satisfying $x_1 \geq \sqrt{x_2^2 + x_3^2}$.}
\label{fig:qcone}
\end{figure}

We define an $n$-dimensional \emph{quadratic cone} as a subset of $\mathbb{R}^n$,
\begin{equation}\label{quadCone}
\Q^n = \left\{ \bfx \in \mathbb{R}^n \mid x_1 \geq \sqrt{x_2^2 + x_3^2 + \cdots + x_n^2}\right\}.
\end{equation}
The geometric interpretation of a quadratic (or second-order) cone is shown in Fig.~\ref{fig:qcone} for a
cone with three variables, and illustrates how the exterior of the cone resembles an ice-cream
cone. 
A convex set $S$ is called a convex cone if for any $\bfx \in S$ we have $\alpha \bfx \in  S\; \forall \alpha \geq  0$. From the definition (\ref{quadCone}) it is clear that if $\bfx \in \Q_n$
then obviously $\alpha \bfx \in \Q_n\; \forall \alpha \geq 0$, which justifies the notion quadratic cone.

An $n$-dimensional rotated quadratic cone is defined as
\begin{equation}\label{rotCone}
\Qr^n = \left\{ \bfx \in \mathbb{R}^n \mid 2x_1x_2 \geq x_3^2 + \cdots + x_n^2,\,x_1,\,x_2\geq0\right\}.
\end{equation}
The two cones are equivalent under an orthogonal transformation, so we only need the first one, but having both is convenient and results in simpler formulations.

\subsection{From quadratic to conic optimisation}
Often regression problems are solved using quadratic optimisation. In Equation (\ref{leastSquare2}) we compute explicitly the inner product of $\bfX \bfw - \bfy$,
\[
\min_{\bfw \in \R^m}\; \bfw^{\T} \bfX^{\T} \bfX \bfw - 2 \bfy^{\T} \bfX \bfw + \bfy^{\T} \bfy.
\]
Any convex quadratic programming problem can be reformulated as a conic programming problem, but the latter class of optimization problems is much more general and yields a more flexible modelling tool. Some simple examples include:
\begin{itemize}
\item $\abs{x} \leq t \iff (t,x) \in \Q^2$. 
\item $\norm{\bfA\bfx - \bfb}_2 \leq t \iff (t, \bfA \bfx- \bfb) \in \Q^{n+1}$
\item $\abs{x}^2 \leq t \iff (1/2, t, x) \in \Qr^3$
\item $\norm{\bfA \bfx-\bfb}_2^2 \leq t \iff (1/2, t, \bfA \bfx-\bfb) \in \Qr^{n+2}$
\end{itemize}
More examples can be found in the MOSEK modeling guide \cite{Letter}.

\subsection{Regression using cones}
A core idea of modern optimisation is to lift a problem in space of a higher dimension in which it has a standard structure much more convenient for modern solvers. This strategy applies to regression problems, too. This may seem counterintuitive but opens a range of new possibilities.

We use the epigraph formulation (e.g., minimizing $f(x)$ is equivalent to minimize $v$ such that $f(x) \leq v$) for Equation (\ref{leastSquare}) to establish the existence of a cone,
\[
\begin{array}{ll}
\min_{(\bfw, v) \in \R^{m+1}} & v\\
\st& \norm{\bfX \bfw-\bfy}_2\leq v.
\end{array}
\]
Note that the constraint describes a quadratic cone, e.g., 
$(v,\,\bfX \bfw-\bfy) \in \Q_{n+1}$. 
For Equation (\ref{leastSquare2}) we use a rotated cone,
\[
\begin{array}{ll}
\min_{(\bfw, v) \in \R^{m+1}} &v\\
\st &\norm{\bfX \bfw-\bfy}^2_2\leq 2\times\frac{1}{2} \times v.
\end{array}
\]
And hence $(1/2,v,\bfX \bfw-\bfy) \in \Qr^{n+2}$.

\subsection{MOSEK}
MOSEK is commercial optimisation solver for large-scale convex and integer conic optimisation. 
The solver implements the \emph{homogeneous embedding} algorithm \cite{YeToMi:94}, which has proven to be 
very robust and reliable. For example, it handles infeasible models in a graceful manner, providing either an
optimal solution or a \emph{certificate} that the problem is infeasible or unbounded. The different cones supported by 
the conic solver in MOSEK are
\begin{itemize}
\item The whole $\mathbb{R}^n$.
\item The nonnegative orthant $\mathbb{R}^n_+$.
\item Quadratic cones.
\item Rotated quadratic cones.
\item The cone symmetric positive semidefinite matrices.
\end{itemize}
Semidefinite cones add significant flexibility and allows modeling of a vast class of problems, 
see \cite{BenTNem:01,BoVa:04},
but is outside the scope of this paper.

\subsection{An example}
In a brief intermezzo we give a first function using Python and the new Fusion interface of MOSEK
to implement the problem
\[
\begin{array}{ll}
\min_{(\bfw, v) \in \R^{m+1}}& v\\
\st& (v,\,\bfX \bfw-\bfy) \in \Q_{n+1}\\
&\sum_{i=1}^m w_i = 1\\
&\bfw \geq 0.
\end{array}
\]
\begin{lstlisting}[caption={Constrained regression}, label=Fragment1]
from mosek.fusion import *

def __rotQCone__(model,expr1,expr2,expr3):
    model.constraint(Expr.vstack(expr1,expr2,expr3), 
                     Domain.inRotatedQCone())

def __lsq__(model,name,X,w,y):
    # append the variable v to the model
    v = model.variable(name, 1, Domain.unbounded())    
    # (1/2, v, Xw-y) \in Qr
    residual = Expr.sub(Expr.mul(DenseMatrix(X),w),y)
    __rotQCone__(M,0.5,v,residual)
    return v
     
def lsqPosFullInv(X, y):
    # define a model
    M = Model('lsqPos')  
    
    # weight-variables
    w = M.variable('w', X.shape[1], Domain.greaterThan(0.0))                      
    
    # e'*w = 1
    M.constraint(Expr.sum(w), Domain.equalsTo(1.0))    
    
    # variable for sum of squared residuals
    v = __lsq__(M,'ssqr',X,w,y)
    
    model.objective(ObjectiveSense.Minimize, v)
    model.solve()
    
    return w.level()
\end{lstlisting}

Fusion is a new high-level interface for Mosek. We construct a model and append variables and constraints on the fly. Please note that the model is passed by reference, e.g. adding a variable in the {\tt lsq} function will modify the model used in the call.

\section{Regularisation}\label{sec:regularisation}
If the columns of $\bfX$ are nearly dependent (i.e., high correlations between the explanatory variables), 
regularisation stabilises the computational results, which are otherwise unreliable and highly sensitive to 
small perturbations and rounding errors.
Such effects are responsible for the bad reputation of (quantitative) portfolio optimisation amongst some practioners, compare with \cite{Michaud}.
In practice it has often been observed that the optimal portfolio takes extreme leverage and alternate positions dramatically when input data is modified.
Regularisation can help to tame an optimiser.


Practioners should be alarmed by potential instabilities in their portfolio optimisation process. Regularisation is often combined with constraints that rule out certain solution, e.g.~further below we will discuss how to control leverage in an equity portfolio. A combination of both and common sense works best in practice. Here we shall explain how to incorporate regularisation terms in least squares problems.

In the context of portfolio optimisation we can interpret regularisation terms as trading costs. Often we reoptimise portfolios when new data is available. The current state of the portfolio is described by a vector $\bfw_0$. Trades $\Delta \bfw$ are induced by changes in the coefficients $w$
\[
\Delta \bfw = \bfw - \bfw_0.
\]
Obviously we want to avoid rather abrupt and dramatic changes as they result in large costs.


There are several potential choices for a trading cost model:
\begin{itemize}
\item Quadratic costs, e.g. $costs \sim  \Delta \bfw^2$. This choice avoids large trades and tends to overestimate trading costs for large trades.
Known as Ridge regression or Tikhonov regularisation.
\item Linear costs, e.g. $costs \sim  \abs{\Delta\bfw}$. This choice is giving preference to sparse updates in $w$ but does not match the nonlinear effects of eating into an order book. Known as sparse regression or LASSO.
\item Subquadratic costs, e.g.~$costs \sim  \abs{\Delta \bfw}^{3/2}$. This choice is motivated by empiric density distributions of the order book, compare with \cite{Almgren}. 
\end{itemize}
The combination of such models is possible. Combining quadratic and linear costs is known as an Elastic Net in modern statistics.

\subsection{Ridge regression}
The regularisation term is included in this minimisation,
\[
\min_{\bfw \in \R^{m}}\; \norm{\bfX\bfw-\bfy}^2_2+\lambda \norm{\Gamma (\bfw-\bfw_0)}^2_2
\]
for some suitably chosen matrix, $\Gamma$. In many cases, this matrix is chosen as the identity matrix $\Gamma= \bfI$, giving preference to solutions with smaller norms.
For the unconstrained problem the modified normal equations,
\[
\left(\bfX^{\T} \bfX + \lambda \Gamma^{\T} \Gamma \right)\bfw =\bfX^{\T} \bfy + \lambda \Gamma^{\T} \Gamma \bfw_0
\]
reveal the proximity to shrinkage for the scaled covariance matrix $\bfX^{\T} \bfX$.  
In the constrained case, a closed-form solution is once again usually not available.

We solve the problem by introducing an additional rotated quadratic cone
\[
\begin{array}{ll}
\min_{(\bfw, v, u) \in \R^{m+2}}& v+u\\
\st&\left(v,\frac{1}{2},\bfX \bfw-\bfr\right) \in \Qr^{n+2}\\
&\left(u,\frac{1}{2},\Gamma (\bfw - \bfw_0)\right) \in \Qr^{m+2}.
\end{array}
\]

\subsection{Sparse regression}
The regularisation term is now the $1$-norm rather than the $2$-norm, giving preference to sparse solutions with smaller norms,
\[
\min_{\bfw \in \R^{m}}\; \norm{\bfX\bfw-\bfy}^2_2+\lambda \norm{\Gamma (\bfw-\bfw_0)}_1.
\]
The $1$-norm of a vector $\bfv \in \mathbb{R}^m$ is
\[
\norm{\bfv}_1 = \sum_{i=1}^m \abs{v_i}.
\]

We solve the problem by appending $m$ quadratic cones of dimension $2$,
\[
\begin{array}{ll}
\min_{(\bfw, v, \bft) \in \R^{2m+1}} & v+ \lambda \sum_{i=1}^m t_i\\
\st &\left(v,\frac{1}{2},\bfX \bfw-\bfy\right) \in \Qr^{n+2}\\
&(t_i,[\Gamma \bfw - \Gamma \bfw_0]_i) \in \Q^2,\, i=1,\ldots,m.
\end{array}
\]

We construct a sparse regression using Python and the new Fusion interface of MOSEK
\[
\begin{array}{ll}
\min_{(\bfw, v, \bft) \in \R^{2m+1}}  & v + \lambda \sum_{i=1}^m t_i\\
\st        & \left(v,\frac{1}{2},\bfX \bfw-\bfy\right) \in \Qr^{n+2}\\
           & (t_i,[\Gamma \bfw - \Gamma \bfw_0]_i) \in \Q^2,\quad i=1,\ldots,m\\
           & \sum_{i=1}^m w_i = 1\\
           & \bfw \geq 0.
\end{array}
\]
    
%
\begin{lstlisting}[caption={Sparse regression}, label=Fragment231]
from mosek.fusion import *

def __QCone__(model,expr1,expr2):
    model.constraint(Expr.vstack(expr1,expr2), 
                     Domain.inQCone())
      
def __abs__(model,name,expr):
    t = model.variable(name, int(expr.size()), 
                       Domain.unbounded())                    
    # (t_i, w_i) \in Q2  or abs(w_i) <= t_i
    for i in range(0, expr.size()):
        __QCone__(model,t.index(i),expr.index(i))
    
    return t

def lsqPosFullInvPenalty(X, y, Gamma, lamb, w0): 
    # define a model
    M = Model('lsqSparse')   
    
    # weight-variables 
    w = M.variable('w', X.shape[1], Domain.greaterThan(0.0)) 

    # e'*w = 1
    M.constraint(Expr.sum(w), Domain.equalsTo(1.0))    
    
    # variable for sum of squared residuals
    v = __lsq__(M,'ssqr',X,w,y)
    
    # variable sum[abs(Gamma*(w-w0))]
    p = Expr.mul(DenseMatrix(Gamma),Expr.sub(w,w0))
    t = Expr.sum(__abs__(M, 'abs(weights)', p))
        
    model.objective(ObjectiveSense.Minimize, 
                    Expr.add(v, Expr.mul(lamb, t)))
    model.solve()
     
    return w.level() 
\end{lstlisting}

\subsection{The $3/2$ regression}
This choice is motivated by empiric investigations of orderbook data. Note that this problem is not a reformulated quadratic problem. It is one of the classic examples revealing the power and flexibility of conic programming,
\[
\min_{\bfw \in \R^m}\;\norm{\bfX\bfw-\bfy}^2_2+\lambda \sum_{i=1}^m \abs{[\Gamma (\bfw-\bfw_0)]_i}^{3/2},
\]
which is equivalent to
\[
\begin{array}{ll}
\min_{(\bfw, v, \bfs, \bft, \bfz) \in \R^{4m+1}} &v+\lambda \sum_{i=1}^m t_i\\
\st&\left(v,\frac{1}{2},\bfX\bfw-\bfy\right) \in \Qr^{n+2}\\
&\left(t_i,[\Gamma \bfw - \Gamma \bfw_0]_i\right) \in \Q^2,\, i=1,\ldots,m\\
&\left(s_i,z_i,t_i\right) \in \Qr^3,\, i=1,\ldots,m\\
&\left(\frac{1}{8},t_i,s_i\right) \in \Qr^3,\, i=1,\ldots,m.
\end{array}
\]
We are now using $3m+1$ cones to describe this problem.

\section{Management of equity portfolios}
We solve common problems in portfolio management using techniques introduced above. 

In all examples discussed below, the matrix $\bfX \in \mathbb{R}^{n \times m}$ describes a time series of $n$ consecutive historic returns of $m$ risky assets, each row index of $\bfX$ corresponding to an investment period, and each column index to an asset. We search for an \emph{optimal} combination of those assets. Different objectives, expectations, constraints, risk appetite, and trading costs, etc. result in a variety of formulations used in practice.

\subsection{Conventions}
An investor wishes to manage a portfolio in $m$ risky
assets. The investor allocates fractions $w_i(t)$ of the risk capital $C$ in asset $i$ over an investment period $[t,t+1]$ (one hour, day, week, month, ...), at
the end of which he/she is prepared to adjust the positions again.

At time $t+1$ the update in $\bfw$ is induced by changes in the underlying input data for the portfolio problem.

The vector $\bfX \bfw = (R_1,R_2,\ldots,R_n)^{\T}$ describes the time series of $n$ consecutive portfolio returns. A portfolio return is the weighted sum of the $m$ linear returns
\[
R_k = \sum_{i=1}^m w_i x_{k,i} = \bfX_{k,\cdot} \bfw
\]
where $x_{k,i}$ is the return of asset $i$ over the $k$-th historic investment period.

\subsection{Minimising the tracking error}
An index tracking portfolio minimises the distance (or tracking error) 
to a given portfolio or index, e.g.~a return time series $r_M$. 
\[
\begin{array}{ll}
\min_{\bfw \in \R^m} &\norm{\bfX \bfw-\bfr_M}^2_2\\
\st & \sum_{i=1}^m w_i = 1\\
& w \geq 0,
\end{array}
\]
which is equivalent to
\[
\begin{array}{ll}
\min_{(\bfw, v) in \R^{m+1}} & v\\
\st&\left(v,\frac{1}{2},\bfX \bfw-\bfr_M\right) \in \Qr^{n+2}\\
&\sum_{i=1}^m w_i = 1\\
&w \geq 0.
\end{array}
\]
We track the portfolio by fully investing the capital $C$ with long positions 
only. Please note that the accumulated returns of both portfolios may differ dramatically. A potential remedy is to use accumulated returns for both the assets and the portfolio or index.

\subsection{Minimising the portfolio variance}
Although one could avoid being exposed to risk by not investing at all, minimum variance portfolios got popular. Such portfolios can be constructed by noting that they track a return time series with $\bfr_M = (0, 0, \ldots, 0)$.
From a theoretical perspective, it should not make sense to invest in such portfolios, but in practice they are observed to perform quite competitively, hence their popularity.

\subsection{Maximising the expected portfolio return}
The best known investment model is the $1$-period Mean-Variance
model of Markowitz. In this model we maximise the expected portfolio return while keeping the estimated risk at or below a predefined level by carefully diversifying across various available assets. 

An investor wishes to actively manage a portfolio in $m$ risky
assets. The investor holds fixed positions $w_i(t)$ in asset $i$ over
an investment period $[t,t+1]$ (one hour, day, week, month, ...), at
the end of which he/she is prepared to adjust the positions again.

The expected return of asset $i$ is $\E[R_i]$. $R_i$ is the random variable
describing the return per unit position in asset $i$ over the investment period $[t,t+1]$. The expectations need to be replaced by estimates
\[
\mu_i \approx \E[R_i]
\]
This is usually done via methods described in Section $5$ using historical prices and other data available at time $t$. Hence the expected portfolio return is
\[
\sum_i w_i \mu_i = \bfw^{\T} \mu.
\]
Hence we solve this problem
\[
\begin{array}{ll}
\max_{\bfw \in \R^m} & \bfw^{\T} \mu\\
\st & \norm{\bfX \bfw}_2^2 \leq \sig2\\
& \sum_{i=1}^m w_i = 1\\
& w \geq 0.
\end{array}
\]
The problem is more interesting once we remove the long only constraint. The solution will most likely almost explode as the optimiser identifies some dramatic risk-offsetting positions. We have discussed this before in the context of regularisation.

A common approach is to control the leverage directly. This can be done with bounds on the $1$-norm of $\bfw$. For a portfolio with short positions,
\[
\sum_{i=1}^m \abs{w_i} > \sum_{i=1} w_i = 1. 
\]
We control the $1$-norm of $\bfw$ by introducing $m$ cones, e.g., $t_i \geq \abs{w_i}$ and hence
\[
(t_i, w_i) \in \Q^2,\, i=1,\ldots,m.
\]
To construct a popular $130/30$ portfolio\footnote{For such portfolios we refinance a long position of up to $1.3 \times C$ with a short position of up to $-0.3 \times C$.} we set
\[
\sum_{i=1}^m t_i \leq 1.3 + 0.3.
\]
A market neutral investor may prefer to work with weights not inducing a long bias. Such investors often use
\[
\sum_{i=1}^m w_i = 0.0
\]
and
\[
\sum_{i=1}^m t_i = 2.0
\]
to bound the size of the long and the offsetting short position.
For practical portfolios we often need additional constraints on individual coefficients and subsets of assets, e.g., belonging to certain sectors.

\subsection{Robust portfolio optimisation}
In the previous sections we have seen how regularisation can be an effective tool in the 
presence of unreliable data. Alternatively we can design a 
\emph{robust} portfolio that explicitly takes the uncertainty into account.
Suppose that the data is uncertain, but belongs to a simple known uncertainty set. 
We can then form a robust estimator that optimises the \emph{worst-case} realisation of the
model over the simple uncertainty set. For sufficiently simple uncertainty sets this 
amounts to a tractable optimisation problem, that is not much harder to solve than the 
non-robust version.

For example, suppose that the vector $\mu$ of expected returns is uncertain, but known to lie inside an
ellipsoid
\[
{\mathcal E} = \{ y \in \mathbb{R}^n \mid y = Au + \mu^0, \; \| u \|_2 \leq 1 \},
\]
where $A$ is a symmetric matrix with nonnegative eigenvalues and $\mu^0$ is 
the center of the ellipsoid, i.e., 
we characterise $\mathcal E$ by a (known) transformation of the unit-ball. Such uncertaintity sets arise naturally as confidence regions of statistical estimators.

The \emph{worst-case return} is then given by the minimum of $\bfw^T \mu$ over $\mathcal E$, i.e.,
as
\[
\min_{\mu \in {\mathcal E}}\bfw^{\T} \mu = \min_{\| u \|_2\leq 1} \bfw^{\T} (Au + \mu^0) = \bfw^{\T} \mu^0 + \min_{\| u \|_2\leq 1} \bfw^{\T} Au.
\]
When $w^{\T}A \neq 0$ the last term is minimised by choosing $u = -(A \bfw)/\| A \bfw \|_2$, in other words,
\[
\min_{\mu \in {\mathcal E}}\bfw^{\T} \mu = \bfw^{\T} \mu^0 - \| A \bfw \|_2.
\]
A portfolio maximising the worst-case return over $\mathcal E$ can then be computed by solving
\[
\begin{array}{ll}
\max_{(\bfw,t) \in \R^{m+1}} & \bfw^{\T} \mu^0 - t\\
\st & \norm{\bfX \bfw}_2^2 \leq \sig2\\
    & \| A \bfw \|_2 \leq t\\
    & \sum_{i=1}^m w_i = 1\\
    & w \geq 0.
\end{array}
\]

\section{Prediction of asset returns}
In the previous section we have already introduced the concept of an 
expected return for an asset. Quantitative portfolio management 
esssentially relies on the assumption that the common disclaimer of previous returns and their lack of indicative power 
for the future is not true. 
Trading autocorrelations in asset returns is one of the most common quantitative investment strategies.

To simplify the technical discussion, we assume we have a time series of returns $r_1, r_2, \ldots$. Note that practioners tend to use volatility adjusted and hence homoscedastic returns.

The goal is to predict $r_n$ as a linear function of historic data available up to this point in time, i.e., on $r_1,\ldots,r_{n-1}$ and any subsequent returns are assumed to be predicted by the same linear function applied to their own analogous data history (with an appropriate shift in time).

We could setup the system the most unconstrained model
\begin{equation}\label{gen}
\begin{pmatrix}
r_1& r_2 & \cdots & r_{n-1} \\
r_2& r_3 & \cdots & r_{n} \\
\vdots & \vdots & \vdots & \vdots
\end{pmatrix}
\bfw =
\begin{pmatrix}
r_n\\
r_{n+1}\\
\vdots
\end{pmatrix}.
\end{equation}
A common alternative is to use simple linear functions of historic data
\begin{equation}\label{movAv}
\begin{pmatrix}
MA_1(r_{n-1}) & MA_2(r_{n-1}) & \cdots \\
MA_1(r_{n}) & MA_2(r_{n}) & \cdots \\
\vdots & \vdots & \ddots
\end{pmatrix}
\bfw =
\begin{pmatrix}
r_n\\
r_{n+1}\\
\vdots
\end{pmatrix}.
\end{equation}
Moving averages are often chosen as linear functions of historic data, so that (\ref{movAv}) is in fact obtained as a restriction of model (\ref{gen}) to $\bfw$ taking values only in a certain linear subspace. This dimensionality reduction allows for $\bfw$ to be computed more accurately than under the unconstrained model (\ref{gen}), since the constrained model is provided with more data relative to the number of degrees of freedom. Another method to increase the number of data points relative to the number of degrees of freedom is to transform the data obtained from different assets in such a way as to make them all appear on the same scale, so that all the data can be used together. This results in more robust estimators of $\bfw$.

Of course, we are by no means restricted to use moving averages. There are numerous interesting ideas oscillating around. 



Common sense combinded with the regularisation techniques in Section \ref{sec:regularisation} can construct very competitive trading systems driven by data. 

\section{Real-world examples}
In this section we sketch typical tasks in portfolio management. Our ultimate goal here is to demonstrate how MOSEK can help in common research problems.

\subsection{Data}
We download equity data from Yahoo finance using the popular pandas library for Python. For the experiments we use adjusted close prices to reflect stock splits, dividends, etc.
We have selected a universe of our $5$ favorite American companies, with symbols 
shown below.
\begin{center}
\begin{tabular}{ll}
GOOG & GOOGLE\\
GS   & Goldman Sachs\\
AAPL & APPLE\\
IMB  & IBM\\
T    & AT\&T\\
\^{}GSPC & S\&P 500 
\end{tabular}
\end{center}
Once fetched we write the data into a csv file to simplify any further analysis.

\begin{lstlisting}[caption={Reading data using pandas}, label=Fragment1]
import pandas.io.data as web
import datetime as dt
import pandas as pd

def fetchDataFromYahoo(symbols):
    s = dt.datetime(2010,  1,  1)
    e = dt.datetime(2012, 12, 31)
    return pd.DataFrame(
        {symb: web.get_data_yahoo(symb, s, e)["Adj Close"]
         for symb in symbols})

if __name__ == '__main__':
    # fetch individual stocks and the S&P index
    symbols = ["GS", "AAPL", "IBM", "GOOG", "T", "^GSPC"]
    fetchDataFromYahoo(symbols).to_csv("data.csv")
\end{lstlisting}

\subsection{Long only equity portfolios}
A common task in quantitative portfolio management is to find a portfolio that would have been optimal in the past.
Obviously, this may bear limited information for the future but we shall avoid this discussion here. We bypass
how often a manager should update a portfolio and how to model the costs that come with rebalancing a portfolio.
Such questions are relevant in practice but beyond the scope of this paper.

Let us now consider fully invested portfolios without short positions. This translates into non-negative coefficients $w$ in a
linear combination of $5$ stocks.

We are reusing the functions defined above. First we compute the minimum variance of a portfolio. The variance of a portfolio return time series
\[
\bfX \bfw =  (R_1,R_2,\ldots,R_n)
\]
is
\[
\Var \bfX \bfw=\frac{1}{n}\sum(R_i - \bar \bfR)^2.
\]
In all but the very slowest trading quantitative trading strategies, the order of fluctuations of returns is at least one order of magnitude larger than th order of the mean. It is therefore common to make the approximating assumption that 
$\bar \bfR = 0$, so that the last problem resembles again a least squares problem
\[
\Var \bfX \bfw = \frac{1}{n} \sum R_i^2 = \frac{1}{n} \norm{\bfX\bfw}_2^2.
\]
In a second question we compute the set of $5$ weights minimizing the variance of the tracking error
\[
\Var \left(\bfX \bfw - \bfr_M\right) = \frac{1}{n} \norm{\bfX \bfw  - \bfr_M}_2^2
\]
where $\bfr_M$ is the return vector of an index.

We also compute the $1/N$ portfolio for our universe, e.g. applying the same weight to each asset.

The results we observe obviously depend on the learning period and in particular on the selected universe of assets. For our universe and the range of dates (2010,2011,2012) we get the following result for the minimum variance portfolio
\begin{verbatim}
GOOG    0.05
T       0.67
AAPL    0.03
GS      0.00
IBM     0.25
\end{verbatim}
and the index tracker
\begin{verbatim}
GOOG    0.10
T       0.30
AAPL    0.12
GS      0.18
IBM     0.30
\end{verbatim}

Note that the minimum variance portfolios tries to invest heavily into AT\&T but avoids Goldman Sachs altogether.
The index tracker is slightly more balanced in its positions.

We also demonstrate how to apply some simple portfolio diagnostic and report an annualized Sharpe ratio and the observed standard deviation of portfolio returns in this period. The observed annualized Sharpe ratio is
\begin{verbatim}
1/N             0.74
Index           0.51
Min Variance    0.93
Tracking        0.77
\end{verbatim}
For the standard deviations of returns we get
\begin{verbatim}
1/N             0.012
Index           0.012
Min Variance    0.009
Tracking        0.011
\end{verbatim}

\begin{lstlisting}[caption={Computing portfolios}, label=Fragment10]
import pandas as pd
import MosekSolver as ms

def computeReturn(ts):
    ts = ts.dropna()
    return ts.diff() / ts.shift(1)

def lsqPosFull(X, y):
    return pd.Series(index=X.columns, 
                     data=ms.lsqPosFull(X.values, y.values))

def AnnualizedSharpeRatio(ts):
    return 16*ts.mean()/ts.std()

if __name__ == '__main__':
    # load data from csv files
    data = pd.read_csv("data.csv", index_col=0, 
                                   parse_dates=True)

    stocks = data[["GOOG","T","AAPL","IBM","GS"]]
    index = data["^GSPC"]

    retStocks = stocks.apply(computeReturn).fillna(value=0.0)
    retIndex = computeReturn(index).fillna(value=0.0)

    rhsZero = pd.TimeSeries(index=retStocks.index, data=0.0)

    wMin = lsqPosFull(X=retStocks, y=rhsZero)
    wTrack = lsqPosFull(X=retStocks, y=retIndex)

    d = dict()
    d["Minimum Variance"] = (retStocks * wMin).sum(axis=1)
    d["Index"] = retIndex
    d["1/N"] = retStocks.mean(axis=1)
    d["Tracking"] = (retStocks * wTrack).sum(axis=1)
    frame = pd.DataFrame(d)

    # apply some diagnostics
    print frame.apply(AnnualizedSharpeRatio)
    print frame.std()
\end{lstlisting}









\section{Conclusions}
Conic programming provides the flexibility needed to solve challenging regression problems. Such problems arise not only in finance but rather in any quantitative discipline dealing with data. In this paper we discussed in particular portfolio optimisation.

\end{document}